\documentclass[10pt, conference, compsocconf]{IEEEtran}
\ifCLASSINFOpdf
\else
\fi
\usepackage{hyperref}
\usepackage{graphicx}
\usepackage{cite}
\usepackage{textcomp}
\usepackage{xcolor}
\usepackage{listings}
\usepackage{float}
\usepackage{array}
\usepackage{wrapfig}
\usepackage{multirow}
\usepackage{tabu}
\usepackage[super]{nth}
\usepackage{cite}
\usepackage{amsmath,amssymb,amsfonts}
\usepackage{algorithm,algorithmic}
\usepackage{makecell}
\usepackage{textcomp}
\usepackage{xcolor}
\usepackage{url}

\makeatletter

\begin{document}
%
\title{Finding Security Vulnerabilities in Unmanned Aerial Vehicles Using Software Verification}


\author{\IEEEauthorblockN{Omar M. Alhawi}
\IEEEauthorblockA{
 University of Manchester\\
 Manchester, UK\\
 omar.alhawi@manchester.ac.uk}
\and
\IEEEauthorblockN{Mustafa A. Mustafa}
\IEEEauthorblockA{
 University of Manchester, UK \&\\
 imec-COSIC, KU Leuven, Belgium\\
mustafa.mustafa@manchester.ac.uk}
\and 

\IEEEauthorblockN{Lucas C. Cordiro}
\IEEEauthorblockA{
 University of Manchester\\
  Manchester, UK\\
lucas.cordeiro@manchester.ac.uk}
}


%


\maketitle

\begin{abstract}
The proliferation of Unmanned Aerial Vehicles (UAVs) embedded with vulnerable monolithic software has recently raised serious concerns about their security due to concurrency aspects and fragile communication links. However, verifying security in UAV software based on traditional testing remains an open challenge mainly due to scalability and deployment issues. Here we investigate software verification techniques to detect security vulnerabilities in typical UAVs. In particular, we investigate existing software analyzers and verifiers, which implement fuzzing and bounded model checking (BMC) techniques, to detect memory safety and concurrency errors. We also investigate fragility aspects related to the UAV communication link. All UAV components (e.g., position, velocity, and attitude control) heavily depend on the communication link. Our preliminary results show that fuzzing and BMC techniques can detect various software vulnerabilities, which are of particular interest to ensure security in UAVs. We were able to perform successful cyber-attacks via penetration testing against the UAV both connection and software system. As a result, we demonstrate real cyber-threats with the possibility of exploiting further security vulnerabilities in real-world UAV software in the foreseeable future. 
\end{abstract}


%
\IEEEpeerreviewmaketitle

\section{Introduction}

Unmanned Aerial Vehicles (UAVs), also sometimes referred to as \textit{drones}, are aircrafts without human pilots on board; they are typically controlled remotely and autonomously and have been applied to different domains (e.g., industrial, military, and education). In $2018$, PWC estimated the impact of UAVs on the UK economy, highlighting that they are becoming essential devices in various aspects of life and work in the UK. The application of UAVs to different domains are leading to GBP $42$bn increase in the UK's gross domestic product and $628,000$ jobs in its economy~\cite{pwc2018}. 
 
With this ever-growing interest also comes an increasing danger of cyber-attacks, which can pose high safety risks to large airplanes and ground installations, as witnessed at the Gatwick airport in the UK in late $2018$, when unknown UAVs flying close to the runways caused disruption and cancellation of hundreds of flights due to safety concerns~\cite{GatwickDronesAttack}. Recent studies conducted by the Civil Aviation Authority show that a $2$kg UAV can cause a critical damage to a passenger jet windscreen~\cite{AviationHouse2018}. Therefore, it remains an open question whether the  {\it Confidentiality}, {\it Integrity}, and {\it Availability} (CIA) triad principles,  which is a model designed to guide policies for information security~\cite{key:article}, will be maintained during UAVs software development life-cycle. 

UAVs typically demand high-quality software to meet their target system's requirements. Any failures in embedded (critical) software, such as those embedded in avionics, might lead to catastrophic consequences in the real world. As a result, software testing and verification techniques are essential 
ingredients for developing systems with high \textit{dependability} and \textit{reliability} requirements, needed to guarantee both user requirements and system behavior. 

Bounded Model Checking (BMC) was introduced nearly two decades ago as a verification technique to refute safety properties in hardware~\cite{handbook09}. However, BMC has only relatively recently been made practical, as a result of significant advances in Boolean Satisfiability (SAT) and Satisfiability Modulo Theories (SMT)~\cite{handbook09}. Nonetheless, the impact of this technique is still limited in practice, due to the current \textit{size} (e.g., number of lines of source code) and \textit{complexity} (e.g., loops and recursions) of software systems. For instance, when a BMC-based verifier symbolically executes a program, it encodes all its possible execution paths into one single SMT formula, which results in a large number of constraints that need to be checked. Although BMC techniques are effective in refuting properties, they still suffer from the state-space explosion problem~\cite{Gadelha2018}.

Fuzzing is a successful testing technique that can create a substantial amount of random data to discover security vulnerabilities in real-world software~\cite{MillerCM06}. However, subtle bugs in UAVs might still go unnoticed due to the large state-space exploration, as recently reported by Chaves et al.~\cite{chaves2018}. Additionally, according to Alhawi et al.~\cite{Alhawi2019}, fuzzing could take a significant amount of time and effort to be completed during the testing phase of the software development life-cycle in addition to its code coverage issues. Apart from these limitations, fuzzing and BMC can enable a wide range of verification techniques. Some examples include automatic detection of bugs and security vulnerabilities, recovery of corrupt documents, patch generation, and automatic debugging. These techniques have been industrially adopted by large companies, including but not limited to Amazon Web Service (CBMC~\cite{CookKKTTT18}), Microsoft (SAGE~\cite{Godefroid2008}), IBM (Apollo~\cite{Artzi2008}), 
and NASA (Symbolic PathFinder~\cite{PAsAreanu}). For example, the SAGE fuzzer has already discovered more than $30$ new bugs in large shipped Windows applications~\cite{Godefroid2008}. Nonetheless, an open research question consists of whether these techniques can be useful in terms of correctness and performance to verify UAV applications. 

Our research investigates both fuzzing and BMC techniques to detect security vulnerabilities in real-world UAV software automatically. Thus, our main research goal is to allow the development of software systems, which are immune to cyber-attacks and thus ultimately improve software reliability. According to the current cyber-attacks profile concerning advanced UAVs, it becomes clear that the current civilian UAVs in the market are insecure even from simple cyber-attacks. To show this point of view, we highlight in our study real cyber-threats of UAVs by performing successful cyber-attacks against different UAV models. These cyber-attacks led to gain a full unauthorized control or cause the UAVs to crash. We show that pre-knowledge of the receptiveness of the UAV system components is all what attackers need to know during their reconnaissance phase before exploiting UAV weaknesses.

\subsection{Contributions}
Our main contribution is to propose a novel approach for 
detecting and exploiting security vulnerabilities in UAVs. We leverage the benefit of using both fuzzing and BMC techniques to detect security vulnerabilities hidden deep in the software state-space. In particular, we make three significant contributions: 
\begin{itemize}
\item Provide a novel verification approach that combines fuzzing and BMC techniques to detect software vulnerabilities in UAV software.
\item Identify different security vulnerabilities that UAVs can be susceptible to. We perform real cyber-attacks against different UAV models to highlight their cyber-threats.
\item Evaluate a preliminary verification approach called ``UAV fuzzer'' to be compatible with the type of UAV software used in industry to exploit their vulnerabilities. 
\end{itemize}

Although our current work represents an ongoing research, preliminary results show that fuzzing and BMC techniques can detect various software vulnerabilities, which are of particular interest to UAV security. We are also able to perform successful cyber-attacks via penetration testing against the UAV both connection and software system. As a result, we demonstrate real cyber-threats with the possibility of exploiting further security vulnerabilities in real-world 
UAV software in the foreseeable future. 

\subsection{Organisation}

The rest of the paper is organised as follows. Section~\ref{sec:Background} describes the UAVs structure and the recent cyber attacks, in addition to other approaches used to verify security in UAVs. Section~\ref{sec:Method} introduces UAV software, UAV communication layer and the methodology to verify it using Fuzzing and Bounded Model Checking techniques. Section~\ref{sec:Status} then describes the benchmarks used and present the results to determine the effectiveness of our approach. Finally, Section~\ref{sec:conslusions} presents our conclusions.

\section{Background}
\label{sec:Background}

\subsection{Generic Model of UAV Systems}
\begin{figure}
\center{\includegraphics[width=85mm]
{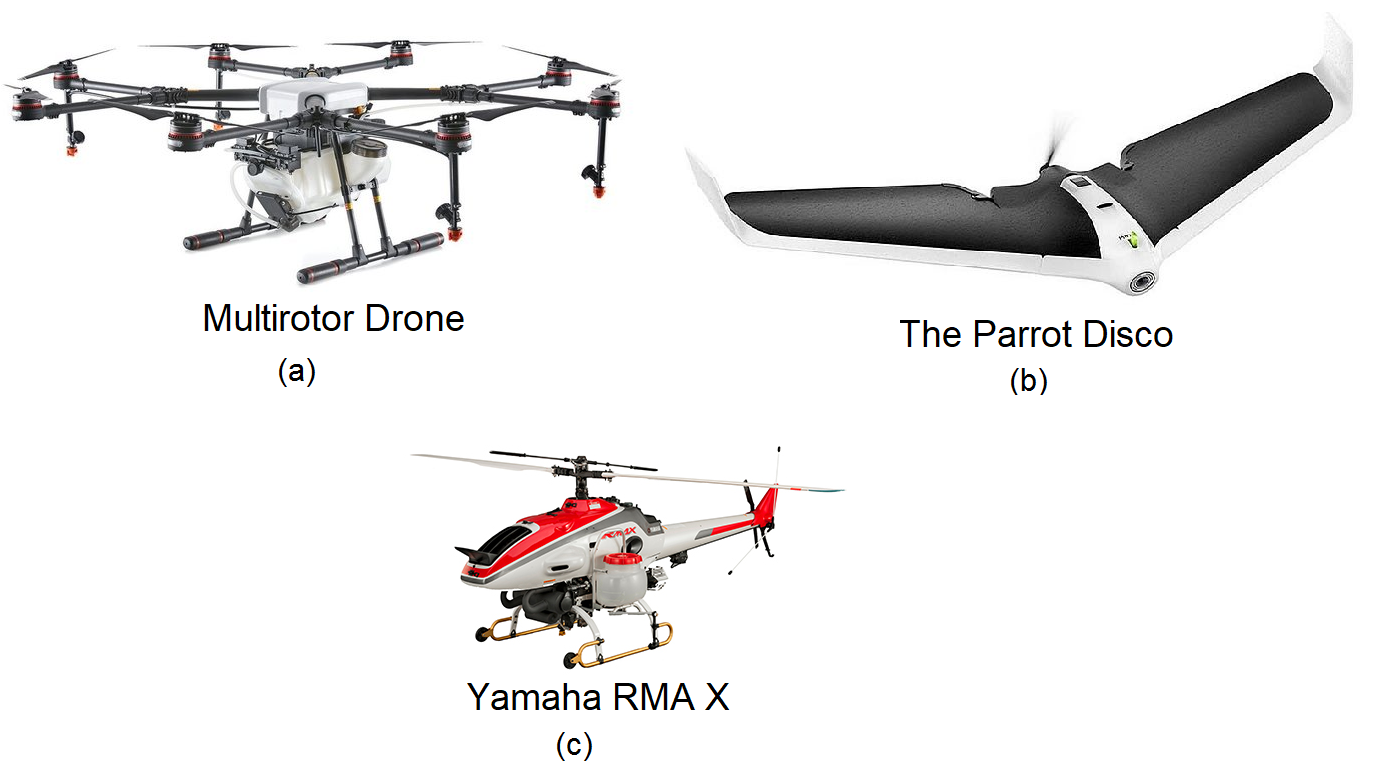}}
\caption{UAV types: multi-rotor (a), fixed wing (b), and single-rotor (c).}
\label{fig:1}
\end{figure}

\begin{figure}[t]
\center{\includegraphics[width=85mm]
{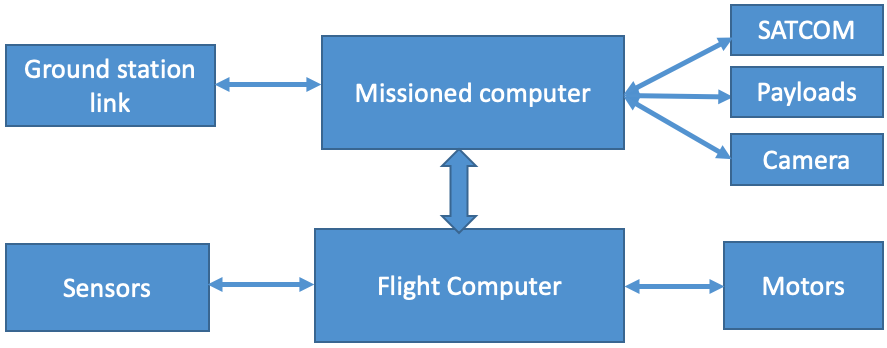}}
\caption{Functional structure of UAVs.}
\label{fig:2}
\end{figure}

Reg Austin~\cite{Austin2010} defines UAVs as a system comprising a number of subsystems, including the aircraft (often referred to as a UAV or unmanned air vehicle), its payloads, the Ground Control Station (GCS) (and, often, other remote stations), aircraft launch and recovery subsystems, where applicable, support subsystems, communication subsystems, and transport subsystems. UAVs have different shapes and models to meet the various tasks assigned to them, such as fixed-wing, single rotor, and multi-rotor, as illustrated in Fig.~\ref{fig:1}. 
However, their functional structure has a fixed standard, as shown in Fig.~\ref{fig:2}. Therefore, finding a security vulnerability in one model might lead to exploiting the same vulnerability in a wide range of different systems~\cite{DeyPCE18,Frei:2006:LVA:1162666.1162671}.

\subsection{Cyber-Threats}

A cyber-threat in UAVs represents a malicious action 
by an attacker with the goal of damaging the surrounding 
environment or causing financial losses, where the UAV 
is typically deployed~\cite{6459914}. In particular, 
with some of these UAVs available to the general public, 
ensuring their secure operation still remains an open research 
question, especially when considering the sensitivity 
of previous cyber-attacks described in the literature~\cite{JoannaFrew2018,JamieTarabay2018}. 

One notable example is the control of deadly weapons 
as with the US military RQ-170 Sentinel stealth aircraft;
it was intercepted and brought down by the Iranian forces 
in late $2011$ during one of the US military operations over 
the Iranian territory~\cite{JoannaFrew2018}. In $2018$, Israel 
released footage for one of its helicopters shooting down 
an Iranian replica model of the US hijacked drone~\cite{JamieTarabay2018}.
Further interest in UAV cyber-security has been raised following this attack. 
For example, Nils Rodday~\cite{Paques2015}, a cyber-security analyst, 
was able to hack UAVs utilized by the police using a man-in-the-middle attack 
by injecting control commands to interact with the UAV.
As a result of previous attacks, UAVs can be a dangerous weapon in the wrong hands. 
Obviously, cyber-attack threats exceeded the cyber-space barrier 
as observed by Tarabay, Lee and Frew~\cite{JoannaFrew2018,JamieTarabay2018}. 
Therefore, enhancing the security and resilience of UAV software 
has become a vital homeland security mission, mainly due to the 
possible damage of cyber-attacks from the deployed UAVs.

\subsection{Verification of Security in UAVs}
\label{sec:RelatedWork}

The UAV components are typically connected together to enable 
secure and fast communication; if one component fails, 
the entire system can be susceptible to malicious attacks~\cite{Austin2010}. 
In this respect, various approaches have been taken 
to automatically verify the correctness of UAVs software. 
In particular, following the RQ-$170$ UAV accident in $2011$, where Iran 
claimed hacking the sophisticated U.S. UAV~\cite{doi:10.1002/rob.21513}, 
a group of researchers from the University of Texas proposed an usual 
exercise to the U.S. Department of Homeland Security; specifically simulated GPS 
signals were transmitted over the air from $620$\,m, 
where the spoofer induced the capture GPS receiver to produce position 
and velocity solutions, which falsely indicated the UAV position~\cite{Bhatti2011}.
A similar study conducted in early $2018$ performed a successful side-channel attack 
to leverage physical stimuli~\cite{Nassi}. The authors were able to detect in 
real-time whether the UAV's camera is directed towards a target or not, by 
analyzing the encrypted communication channel, which was transmitted from a real UAV. 
These prior studies were able to highlight GPS weaknesses, 
but they did not cover the UAV security issues w.r.t. all involved software 
elements, mainly when zero-day vulnerabilities are associated 
with the respective UAV outputs, i.e., a real UAV software bug that is unknown to the
vendor responsible for patching or otherwise fixing the bug.

Other related studies focus on automated testing~\cite{7152319} and 
model-checking the behavior of UAV systems~\cite{chaves2018,Sirigineedi2009}. 
For example, a verification tool named as Digital 
System Verifier (DSVerifier)~\cite{chaves2018} formally checks 
digital-system implementation issues, in order to investigate 
problems that emerge in digital control software 
designed for UAV attitude systems (i.e., software errors 
caused by finite word-length effects). Similar work also 
focuses on low-level implementation aspects, where 
Sirigineedi et al.~\cite{Sirigineedi2009} applied a formal 
modeling language called SMV to multiple-UAV missions 
by means of Kripke structures and formal verification 
of some of the mission properties typically expressed 
in Computational Tree Logic. In this particular study, 
a deadlock has been found and the trace generated by SMV 
has been successfully simulated. Note that these 
prior studies concentrate mainly on the low-level implementation 
aspects of how UAVs execute pilot commands. By contrast, 
we focus our approach on the high-level application 
of UAVs software, which is typically hosted by the firmware embedded in UAVs.

Despite the previously discussed limitations, 
BMC and Fuzzing techniques have been successfully used to verify the 
correctness of digital circuits, security, and communication protocols~\cite{Sirigineedi2009,domin2016security}. 
However, given the current knowledge in ensuring security of UAVs, 
the combination of fuzzing and BMC techniques have not been used 
before for detecting security vulnerabilities in UAV software
(e.g., buffer overflow, dereferencing of null pointers, and 
pointers pointing to unallocated memory regions).
UAV software is used for mapping, aerial analysis and to get optimized images. 
In this study, we propose to use both techniques to detect 
security vulnerabilities in real-world UAV software.

\section{Finding Software Vulnerabilities in UAVs Using Software Verification}
\label{sec:Method}

\subsection{Software In-The-Loop}

UAV software has a crucial role to operate, manage, and 
provide a programmatic access to the connected UAV system. 
In particular, before a given UAV starts its mission, 
the missioned computer, as illustrated in Fig.~\ref{fig:2}, 
exports data required for this mission from a computer running 
the flight planning software. Then, the flight planning software 
allows the operator to set the required flight zone 
(way-point mission engine), where the UAV will follow this 
route throughout its mission instead of using a traditional 
remote controller directly~\cite{7152319}. 

Dronekit\footnote{\url{https://github.com/dronekit/dronekit-python}} 
is an open-source software project, which allows one to command 
a UAV using Python programming language~\cite{van1995python}; 
it enables the pilot to manage and direct control over the UAV movement and operation, 
as illustrated in Fig.~\ref{figDFIcode0}, where one can connect 
to the UAV via a User Datagram Protocol (UDP) endpoint (line~\ref{python:connect})
with the goal of gaining control of the UAV by means 
of the ``vehicle'' object. In particular, UDP allows establishing 
a low-latency and loss-tolerating connection between the pilot and the UAV.
This control process relies 
on the planning software inside the UAVs system, 
which in some cases the software might be 
permanently connected to the pilot controlling system 
(e.g., Remote Controller or GCS) 
due to live feedback or for real-time streaming.

\lstset{language=python,escapechar=|,basicstyle=\small\ttfamily}
\lstset{numbers=left, numberstyle=\tiny, stepnumber=1, numbersep=5pt}
\lstset{tabsize=2}
\lstset{firstnumber=1}
\lstset{frame=single}
\begin{figure}
\centering
\begin{minipage}{\columnwidth}
\tiny	
\begin{lstlisting}[xleftmargin=.025\textwidth,frame=single, basicstyle=\scriptsize]
from dronekit import connect 
# Connect to UDP endpoint.
vehicle = connect('127.0.0.1:14550', wait_ready=True) |\label{python:connect}|
# Use returned Vehicle object to query device state:
print("Mode: %s" % vehicle.mode.name)
\end{lstlisting}
\end{minipage}
\caption{Python script to connect to a vehicle (real or simulated).} 
\label{figDFIcode0}
\end{figure}

Our main research goal is to investigate in depth open-source UAVs code (e.g., 
DJI Tello\footnote{\url{https://github.com/dji-sdk/Tello-Python} }
and Parrot Bebop\footnote{\url{https://github.com/amymcgovern/pyparrot}}) to search for potential security 
vulnerabilities. For example, Fig.~\ref{figDFIcode} shows a simple Python 
code to read and view various data status of Tello UAV. 
In particular, this Python code imports and defines the 
required libraries (lines from~\ref{python:import1} to~\ref{python:import3}) 
and then connects the GCS to the UAV by using the 
predefined port and IP address in lines~\ref{python:tello_ip} and~\ref{python:tello_port}. 
As we can see from lines~\ref{python:trybegin} to~\ref{python:tryend}, 
the UAV will acknowledge the pilot commands and print the Tello current status. If an attacker 
is able to scan and locate the IP address that this particular UAV 
has used, then he/she would be able to easily intercept 
the data transmitted, inject a malicious code or 
take the drone out of service using a denial of service attack,
which can lead the UAV to a crash, thus making it inaccessible. 
In order to detect potential security vulnerabilities in UAV software, 
we provide here an initial insight of how to combine BMC and 
fuzzing techniques with the goal of exploring the system state-space
to ensure safe and secure operations of UAVs.


\lstset{language=python,escapechar=|,basicstyle=\small\ttfamily}
\lstset{numbers=left, numberstyle=\tiny, stepnumber=1, numbersep=5pt}
\lstset{tabsize=2}
\lstset{firstnumber=1}
\lstset{frame=single}
\begin{figure}
\centering
\begin{minipage}{\columnwidth}
\tiny	
\begin{lstlisting}[xleftmargin=.025\textwidth,frame=single, basicstyle=\scriptsize]
import socket  |\label{python:import1}|\label{python:import11}
from time import sleep |\label{python:import2}|
import curses |\label{python:import3}|
INTERVAL = 0.2
...
local_ip = ''
local_port = 8890
socket=socket.socket(socket.AF_INET,socket.SOCK_DGRAM) |\label{python:udp}|
# socket for sending cmd
socket.bind((local_ip, local_port))
tello_ip = '192.168.10.1' |\label{python:tello_ip}|
tello_port = 8889 |\label{python:tello_port}|
tello_adderss = (tello_ip, tello_port)
socket.sendto('command'.encode('utf-8'), tello_adderss)
try: |\label{python:trybegin}|
  index = 0
  while True:
    index += 1
    response, ip = socket.recvfrom(1024)
    if response == 'ok':
      continue
    out = response.replace(';', ';\n')
    out = 'Tello State:\n' + out
    report(out)
    sleep(INTERVAL) |\label{python:tryend}|
...
\end{lstlisting}
\end{minipage}
\caption{Python code fragment to read and view various data status of Tello UAV.} 
\label{figDFIcode}
\end{figure}


\subsubsection{Illustrative Example Using UAV swarm}

\begin{figure}
\center{\includegraphics[width=90mm]
{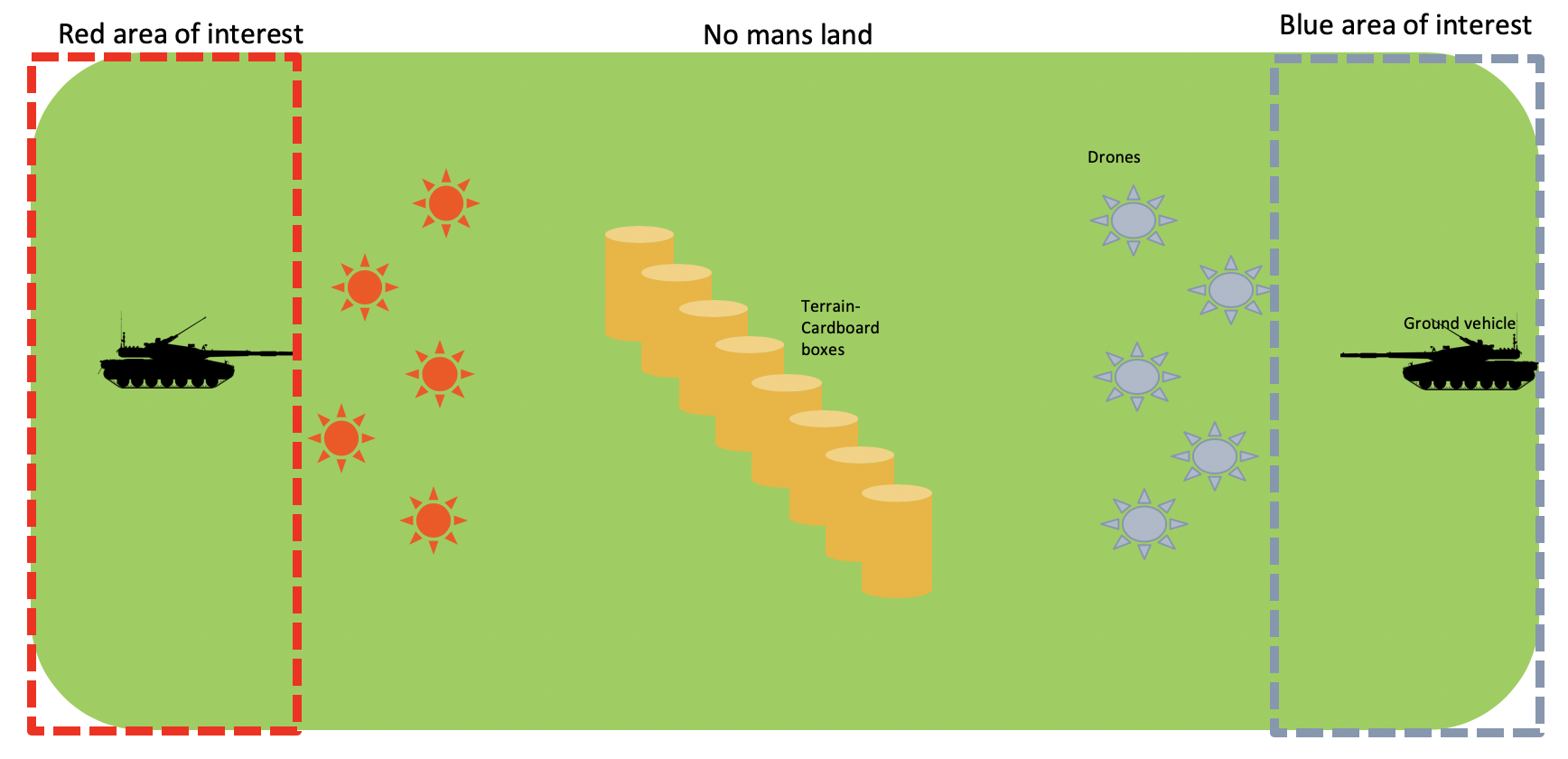}}
\caption{UAV Swarm Competition.}
\label{fig:4}
\end{figure}

Throughout this paper, we use an illustrative example from UAV swarm, 
which consists of multiple UAVs to autonomously make decisions based 
on shared information; the safe and secure operation of multiple UAVs is 
particularly relevant since they have the potential to revolutionize the dynamics of conflict.
In early $2019$, we participated in a competitive-exercise\footnote{\url{https://www.youtube.com/watch?v=dyyaY1VXqL4}}, 
with five different UK universities; the main goal of this event consisted of teams 
from across the UK to compete against each other in a game of offense (\textcolor{red}{red} team) 
and defense (\textcolor{blue}{blue} team) using swarms of UAVs, as illustrated in Fig.~\ref{fig:4}. As a result,
this competition allowed us to highlight aspects of how to protect 
urban spaces from UAV swarms, which is a serious concern of modern society.
This competition was sponsored by the British multinational defense, 
security, and aerospace company (BAE). Solutions developed 
by industry, such as the ``jamming guns'' and single ``UAV catchers'', 
fall short of what would be required to defend against 
a large automated UAV swarm attack. 
For this particular illustrative example, using software verification 
and the UAV connection weakness, we were able to perform 
a successful cyber-attack against UAV models by scanning 
the radio frequencies and targeting the unwanted UAVs 
with just a raspberry-pi, a Linux OS installed on and $2.4$ \textit{GHz} 
antennas, as reported in our experimental evaluation. 




\subsection{Verifying UAV Software using Fuzzing and BMC}
\label{FuzzingBMCApproach}

We describe our novel verification approach approach called ``UAV Fuzzer'' 
to check for security vulnerabilities in UAVs. In particular, we check
for user-specified assertions, buffer overflow, memory safety, division by zero, 
and arithmetic under- and overflow. Our verification approach consists of running 
a fuzzer engine using pre-collected test cases $TC=\{tc_1, tc_2, \ldots, tc_n\}$, 
where $n$ represents the total number of pre-collected test cases, with the goal of 
initially exploring the state-space of the UAV software operation. Note that a test case $tc_i$ 
used by our approach is similar to a real valid data, but it must contain a problem on it, 
or also called ``anomalies''. For example, to fuzz UAV software, a test case should be a connection between the UAV and GCS, 
so the mutated version generated of such a similar connection is called a test case $tc$. 

In our ``UAV Fuzzer'', we keep track of each computation path of the program $\Pi = \{\pi_1, \pi_2, \ldots, \pi_n \}$, 
which represents the program unfolding for a bound $k$ and a security property $\phi$ initially 
explored by our fuzzer. If our fuzzer engine gets stuck due to the mutations generated are 
not suited enough to the new state transition, our BMC tool runs against the target software 
to symbolically explore its uncovered state-space with the goal 
of checking the unexplored execution paths in $\Pi$ of the UAV software.
The idea behind BMC is to check the negation of a given property $\phi$ at a given depth $k$, 
i.e., given a transition system \textit{M}, a property $ \phi $, and a limit of iterations \textit{k}, 
BMC unfolds a given system \textit{k} times and converts it into a Verification Condition (VC) 
$\psi$, such that $\psi$ is \textit{satisfiable} if and only if $\phi$ has a counterexample ($cex$)
of depth less than or equal to \textit{k}. 


We formally describe our verification algorithms ``UAV Fuzzer'' by assuming that a given
program $P$ under verification is a state transition system $M$.
In $M$, a state $s \in S$ consists of the value of the program counter and the
values of all program variables. A predicate $init_P(s)$ denotes that $s$ is an
initial state, $tr_p(s_i, s_j) \in T$ is a transition relation from $s_i$ to
$s_{j}$, $\phi(s)$ is the formula encoding for states satisfying a safety
property, and $\psi(s)$ is the formula encoding for states satisfying a
completeness threshold~\cite{Kroening2011}, which is equal to the maximum number
of loop iterations occurring in $P$. For convenience, we define an error state
$\epsilon$, reachable if there exists a property violation in the program
$P$. A counterexample $cex^{k}$ is a sequence of states of length $k$ from an
initial state $s_1$ to $\epsilon$. The main steps for our proposed verification 
algorithm are described in Algorithm~\ref{alg:verification-algorithm}.
 \begin{algorithm}[h]
 \caption{UAV Fuzzer}
 \begin{algorithmic}[1]
 \renewcommand{\algorithmicrequire}{\textbf{Input: Program}}
 \renewcommand{\algorithmicensure}{\textbf{Output: Verification Successful or Failed}}
  \STATE Define pre-collected test cases $TC=\{tc_1, tc_2, \ldots, tc_n\}$ to be employed by the fuzzing engine. \\
  \STATE Fuzzer engine begins to explore each execution path $\pi_i$ starting from an initial state $s$ and produces malformed inputs  $I = \{ \iota_{1}, \iota_{2}, \ldots, \iota_{n} \}$ to test for potential security vulnerabilities.
  \STATE Store each $\pi_i$ that has been verified and repeat step $2$ until the fuzzer engine either reaches a crashing point or it cannot explore the next $\pi_i$ in $\Pi$ due to complex guard checks. \\
  \STATE Run BMC to verify the remaining execution paths in $\Pi$ that have not been previously explored by our fuzzer engine in steps $2$ and $3$. \\
  \STATE Repeat step $4$ until BMC falsifies or verifies $\phi$ or it exhausts time and memory limits.
  \STATE Once BMC completely verifies the UAV code in step $5$, it returns ``false'' if a property violation is found together with $cex^{k}$, ``true'' if it is able to prove correctness, or ``unknown''.\\
 \end{algorithmic} 
 \label{alg:verification-algorithm}
 \end{algorithm}

As an illustrative example, consider the code fragment shown in Fig.~\ref{figDFIcode}. First, our UAV fuzzer starts by defining new data based on the input expected by our targeted model (Tello UAV). As an example, Fig.~\ref{code:test-case} shows a valid test case generated by our UAV fuzzer, which is based on the  module specification for the Tello UAV; this test case expects the IP address and specific port before launching the drone to start flying.
Second, our UAV fuzzer engine starts exploring and running the UAV software (cf.~Fig.~\ref{figDFIcode}) with the generated test-cases (cf.~Fig.~\ref{code:test-case}) and then it records each execution path $\pi_i$ that has been explored. Third, when the UAV fuzzer engine reaches a complex condition and struggles to find its next path (line~\ref{python:if} of Fig.~\ref{code:test-case}), UAV fuzzer will attempt to reconstruct the following path using BMC, which stores the current fuzzing transactions and restores the next path symbolically. Lastly, BMC will check for any further exception that occurs as a result of its execution. Additionally, UAV fuzzer can report the code coverage achieved during the testing process, and thus provide a better understanding of code coverage status.
\lstset{language=python,escapechar=|,basicstyle=\small\ttfamily}
\begin{figure}[!ht]
\centering
\begin{minipage}{\columnwidth}
\tiny	
\begin{lstlisting}[xleftmargin=.025\textwidth,frame=single, basicstyle=\scriptsize,label={UAV_testcase}]
  while True:
    index %= 1
# + replaced with %
    response, ip = socket.recvfrom(1024)
    if response == 'ok' |\label{python:if}|
      continue
\end{lstlisting}
\end{minipage}
\caption{Test case from the Tello UAV embedded software.}
 \label{code:test-case}
\end{figure}

\subsection{UAV Communication Channel}

An UAV has a radio to enable and facilitate remote communication between 
the GCS and the UAV. In addition, it consists 
of different electronic components, which interact autonomously 
with a goldmine of data transmitted over the air during its flight's  
missions; this makes the communication channel in UAVs 
an ideal target for a remote cyber-threat. Therefore, ensuring
secure (bug-free) software, together with a secure communication channel, 
emerges as a priority in successful deployment of any UAV system. 

A successful false-data injection attack was demonstrated 
by Strohmeier et al.~\cite{10.1007/978-3-319-20550-2_4}, which had 
devastating effects on the UAV system. The authors were able 
to successfully inject valid-looking 
messages, which are well-formed with reasonable data into the 
Automatic Dependent Surveillance-Broadcast (ADS-B) protocol. 
Note that this protocol is currently the only means of air 
traffic surveillance today, where Europe and US must comply 
with the mandatory use of this insecure protocol by the end of $2020$\cite{10.1007/978-3-319-20550-2_4}.

\begin{figure}[t]
  \center{\includegraphics[width=90mm]
  {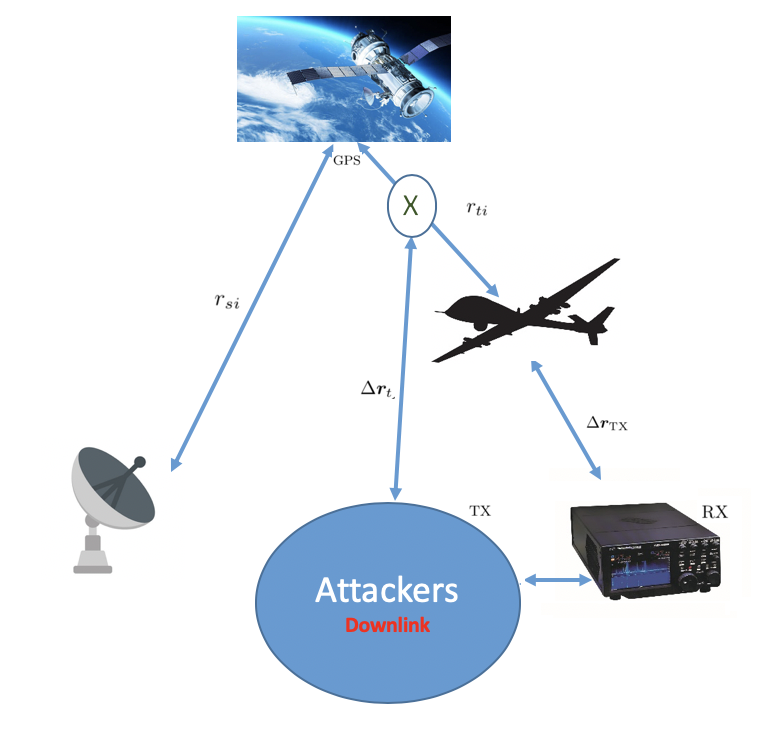}}
  \caption{\label{fig:my-label} GPS-satellite-signal is overlaid by a spoofed GPS-signal.}
\end{figure}

To investigate this layer further, we used 
Software-Defined Radio (SDR) system to receive, 
transmit, and analyze the UAV operational connection 
system (e.g., Ku-Band and WiFi). We have also investigated 
the information exchanged between UAV sensors and the surrounding 
environment for any potential security vulnerabilities 
(e.g., GPS Spoofing), as illustrated in Fig.~\ref{fig:my-label}. 
The signal that comes from the satellite is weak; hence, 
if an attacker uses a local transmitter under the same frequency, 
this signal would be stronger than the original satellite signal. 
As a result, the spoofed GPS-signal will override current satellite-signal, 
thereby leading to spoof a fake position for the UAV targeted. 
In this particular case, the UAV would then be hijacked and put in hold,
waiting for the attacker's next command. Therefore, verifying the 
UAV software to build practical software systems with strong 
safety and security guarantees is highly needed in practice.
      
 
Our GPS spoofing attack is described in Algorithm~\ref{alg:GPSSpoofingAttack}. 
 Here spoofer refers to a Full-Duplex device that is used to attack a particular UAV system to crash or take control of the UAVs. In our experimental evaluation with the UAV models DJI Tello and Parrot Bebop 2, we were able to perform a successful attack under 2.4 {\it GHz} to the target system. First,  we were able to detect the drone frequency by one of the antennae, which configured to monitor the active 2.4 GHz connection referred to as RX. The targeted drones are allocated based on identifying the drone default mac addresses owned by the Parrot company\footnote{http://standards-oui.ieee.org/oui/oui.txt} and their unique SSID. Second, the other antenna was used to transmit the attacker new data referred to as TX. The distance between the attacker equipment and the target (${r}_{si}$ and ${r}_{ti}$) is vital for this attack,  since the antenna/system strength and the delay caused are all taken into consideration during the attack.

 
 
 
  \begin{algorithm}[!h]
 \caption{GPS Spoofing Attack}
 \begin{algorithmic}[1]
 \renewcommand{\algorithmicrequire}{\textbf{Input: Program}}
 \renewcommand{\algorithmicensure}{\textbf{Output: Verification Successful or Failed}}
  \STATE The spoofer device should be located with the nominated antennas (i.e., 2.4 {\it GHz}). \\
  \STATE The antenna configured on monitoring mode ({\it RX}) to detect the authentic signal from the available GPS satellites.
  \STATE The vectors $\Delta$${\bf{r}}_{TX}$ and $\Delta$${\bf{r}}_{t}$ are for the antenna ({\it TX}) coordinates which distributed in a 3-dimensional array. \\
  \STATE ${r}_{si}$ and ${r}_{ti}$ are the respective distances from the GPS satellite.\\
 \end{algorithmic} 
 \label{alg:GPSSpoofingAttack}
 \end{algorithm}
\section{Preliminary Experimental Evaluation}
\label{sec:Status}

We have performed a preliminary evaluation of our proposed 
verification approach to detect security vulnerabilities in UAVs. 
Our proposed method was implemented in a tool called DepthK\cite{depthk}, 
using BMC technique and invariant generators such as PIPS~\cite{pips:2013} and PAGAI~\cite{Henry:2012}. 
PIPS is an inter-procedural source-to-source compiler framework for C and Fortran programs, 
PAGAI is a tool for automatic static analysis that is able to generate inductive invariants, 
both rely on a polyhedral abstraction of program behavior for infering invariants.
We have evaluated the DepthK tool over a set of 
standard C benchmarks, which share common features of UAV code 
(e.g., concurrency and arithmetic operations). We have also evaluated 
our fuzzer engine to test a PDF software with the goal
of checking its efficiency and efficacy to identify bugs. Lastly,
we present our results in the swarm competition promoted by BAE systems.  

\subsection{Description of Benchmarks}

The International Software Verification Competition (SV-COMP)~\cite{Beyer19}, 
where DepthK participated, was run on a Linux Ubuntu $18.04$ OS, 
$15$ GB of RAM memory, a run-time limit of $15$ minutes 
for each verification task and eight processing units of $i7-4790$ CPU. 
The SV-COMP's benchmarks used in this experimental evaluation include:	
%
\textit{ReachSafety}, which contains benchmarks for checking reachability of an error location;	
\textit{MemSafety}, which presents benchmarks for checking memory safety;	
\textit{ConcurrencySafety}, which provides benchmarks for checking concurrency problems;	
\textit{Overflows}, which is composed of benchmarks for checking whether variables of signed-integers type overflow;	
\textit{Termination}, which contains benchmarks for which termination should be decided;	
\textit{SoftwareSystems}, which provides benchmarks from real software systems.	

Our fuzzing experiments were ran on MacBook Pro laptop with 
$2.9$ GHz Intel Core i$7$ processor and $16$ GB of memory. 	
We ran our fuzzing engine for at most of $12$ hours for 
each single binary file. We analyzed and replayed the 
testing result after a crash was reported or after 
the fuzzer hit the time limit. To analyze 
the radio frequencies, we configured/compiled the required software for this purpose 
(e.g. bladerf, GQRX, OsmoSDR, and GNU Radio tool) using bladerf x40 device, 
ALFA high gain USB Wireless adapter and $2.4$ \textit{GHz} antennas.
Additionally, we used the open-source UAVs 
code DJI Tello and Parrot Bebop.

\subsection{Objectives}

The impact of our study is a novel insight on the UAV security potential risks. 
In summary, our evaluation has the following three experimental questions to answer: 
\begin{itemize}
\item[EQ1] {\bf(Localization)} Can DepthK help us understand the security vulnerabilities that have been detected?
\item[EQ2] {\bf(Detection)} Can generational or mutational fuzzers be further developed to detect vulnerabilities in real-world software?
\item[EQ3] {\bf{(Cyber-attacks)}} Are we able to perform successful cyber-attacks in commercial UAVs?
\end{itemize}

\subsection{Results}
\label{sec:results}


\subsubsection{SV-COMP}
\label{sec:svcomp}

Concurrency bugs in UAVs 
are one of the most difficult vulnerabilities to verify~\cite{Sirigineedi2009}. 
Our software verifier DepthK~\cite{depthk} has been used to verify and 
falsify safety properties in C programs, using BMC and 
\textit{k}-induction proof rule techniques. 
In late $2018$, we participated with the DepthK tool in 
SV-COMP 2019\footnote{\url{https://sv-comp.sosy-lab.org/2019/}}
against other software verifiers. Our verifier showed promising 
results over thousands of verification tasks, which are of particular 
interest to UAVs security ({{\it e.g.,} \textit{Concurrency Safety} 
and \textit{Overflows} categories}), which answers \textbf{EQ1}. 

\textit{Concurrency Safety} category, which consists of 
$1082$ benchmarks of concurrency problems, is one of the 
many categories verifiers run over; 
DepthK was able to accurately detect $966$ problems from this category. 
For the \textit{Overflows} category, which consists 
of $359$ benchmarks for different signed-integers overflow bugs, 
DepthK was able to detect $167$ problems. 
These results are summarized in Table~\ref{tab:1}. 
A task counts as \emph{correct true} if it does not contain any reachable 
error location or assertion violation, and the tool reports 
``safe''; however, if the tool reports ``unsafe'', it counts as
\emph{incorrect true}. Similarly, a task counts as 
\emph{correct false} if it does contain a reachable violation, 
and the tool reports ``unsafe'', together with a confirmed
witness (path to failure); otherwise, it counts as
\emph{incorrect false} accordingly.
Dirk Beyer~\cite{Beyer19} 
shows DepthK's results when compared with other verifiers in SV-COMP 2019. 

\begin {table}
\caption {DepthK Results in SV-COMP 2019.} \label{tab:1} 
\begin{tabular}{|p{2.5cm}|p{1cm}|p{1cm}|p{1cm}|p{1cm}|}
 \hline
 {\bf  Category list} & {\bf Correct True} & {\bf Correct False} & {\bf Incorrect Results} & {\bf Unknown} \\
\hline
 Concurrency Safety &  194  &  772  & 20 &  96  \\
  \hline
  Overflows  &  17 &  150 & 0 &  192  \\
  \hline
\end{tabular}
\end {table}



\subsubsection{Fuzzing Approach}
\label{sec:FuzzingApproach}


According to a prior study~\cite{Alhawi2019}, the 
generalizing fuzzing approach leads to a better 
result in discovering and recording software vulnerabilities 
compared with the mutational fuzzing approach 
if the test cases used in the fuzzing experiment 
are taken into account, which answers \textbf{EQ2}. Our experimental results 
applied to a PDF software called Sumatra 
PDF\footnote{\url{https://www.sumatrapdfreader.org/free-pdf-reader.html}}, 
which was chosen for evaluation purposes, are 
shown in Table~\ref{tab:3}. Here, the generational fuzzer was able to detect
$70$ faults in $45$ hours in the Sumatra PDF, while the mutational fuzzer was able to detect
$23$ in $15$ hours.

\begin {table}
\caption {Fuzzing Approaches Comparison.} \label{tab:3} 
\begin{tabular}{|p{2.5cm}|p{2.5cm}|p{1cm}|p{1cm}|}
 \hline
 {\bf Fuzzing Approaches} & {\bf Target} & {\bf Time} & {\bf Faults} \\
 \hline
 Generational Fuzzer & Sumatra PDF  & 45 hours   & 70\\
 \hline
 Mutational Fuzzer & Sumatra PDF  & 15 hours  & 23\\
\hline
\end{tabular}
\end {table}


\subsubsection{UAV Swarm Competition}
\label{sec:UAVSwarmCompetition}


As part of our participation at the UAV swarm competition sponsored by 
(BAE)\footnote{\url{https://www.cranfield.ac.uk/press/news-2019/bae-competition-challenges-students-to-counter-threat--from-uavs}}, penetration testing was performed 
against the UAV both connection and software system, 
in which we were able to perform successful cyber-attacks, which answers \textbf{EQ3}. 
These attacks led to deliberately crash UAVs or to take control 
of different non-encrypted UAV systems (e.g., Tello and Parrot Bebop 2). 
This was achieved by sending connection requests to shut down a UAV CPU, 
thereby sending packets of data that exceed the capacity allocated by the buffer 
of the UAV's flight application and by sending a fake information packet to the device's controller. 
These results are summarized in Table~\ref{tab:2}, where we describe the employed UAV models 
and tools and whether we were able to obtain full control or crash. Note that due to the limitations 
of the competition, DepthK tool was not employed during the BAE competition; however, exploiting potential 
UAV software vulnerabilities is still a continuous research, where we intend to further exploit DepthK.

\begin {table}
\caption {Results of the UAV Swarm Competition.} \label{tab:2} 
\begin{tabular}{|p{2.2cm}|p{1.6cm}|p{1.3cm}|p{1.7cm}|}
 \hline
 {\bf Vulnerability Type} & {\bf UAV Model} & {\bf Tool} & {\bf Result} \\
 \hline
 Spoofing & \center{DJI Tello}  & \center{Wi-Fi transmitter}  & Full Control\\

  Denial of service &  &  &  Full Control \\
 \hline
 Spoofing & \center{Parrot bebop 2}  & \center{Wi-Fi transmitter}  & Full Control\\

 Denial of service &   &   & Crash\\
\hline
\end{tabular}
\end {table}
\subsection{Threats to Validity}

{\it Benchmark selection:} We report the evaluation of 
our approach over a set of real-world benchmarks, where 
the UAVs share the same component structure. Nevertheless, 
this set is limited within our research scope and the experiment 
results may not generalize to other models because other UAV models
have a proprietary source-code. Additionally, we have not evaluated
our verification approach using real UAV code written in Python, 
which is our main goal for future research.

{\it Radio Spectrum:} The frequencies we report on our evaluation 
were between $2.4$ {\it GHz} and $5.8$ {\it GHz}, as the two most 
common ranges for civilian UAVs; however, the radio regulations 
in the UK are complicated (e.g., we are required to be either 
licensed or exempted from licensing for any transmission over the air). 

\section{Conclusions and Future Work}
\label{sec:conslusions}

Our ultimate goal is to develop a UAV fuzzer to be 
introduced as mainstream into the current UAV programming system, 
in order to build practical software systems robust to cyber-attacks.
We have reported here an initial insight of our verification approach
and some preliminary results using similar software typically used by
UAVs. In order to achieve our ultimate goal, we have various tasks 
planned as follows:

\begin{itemize}

\item {\bf Vulnerability Assessment}: Identify and implement 
simple cyber-attacks from a single point of attack against 
different UAV models. We will continue investigating Python 
vulnerabilities at the high-level system (e.g., UAV applications) 
and whether UAVs software is exploitable to those security vulnerabilities. 
\item {\bf Python Fuzzer}: We will develop an automated python 
fuzzer by analyzing how to convert the UAV command packets 
into a fuzzing ones, in order to produce test cases, 
which are amenable to our proposed fuzzer.
\item {\bf GPS Analysis}: We identified based on numerical 
analysis on GPS, the cyber-attack UAVs might be vulnerable from. 
This investigation will continue to develop and simulate 
a GPS attack applied to a real UAV system.
\item {\bf Implementation}: Apply our proposed verification approach 
to test real-world software vulnerabilities, which can be implemented 
during the software development life-cycle to design a cyber-secure architecture.
\item {\bf Evaluation and Application}: Evaluate our proposed approach 
using real-world UAV implementation software. We will also compare 
our approach in different stages to check its effectiveness and efficiency.
\end{itemize}
\section*{Acknowledgment}
Mustafa A.  Mustafa  is  funded  by  the  Dame Kathleen  Ollerenshaw Fellowship awarded by The University of Manchester.



%



\end{document}